This work was supported in part by the National Natural Science Foundation under Grant 62171014, National Natural Science Foundation of China 82201244, Natural Science Foundation of Beijing M22019, Beijing Hospitals Authority Innovation Studio of Young Staff Funding Support 202106 and from the UKRI EPSRC, under grants EP/K03099X/1, EP/S023283/1. *Corresponding authors: Shuo Gao (shuo_gao@buaa.edu.cn) and Luigi G. Occhipinti (lgo23@cam.ac.uk).* Yong Liu and Mengtian Kang contributed equally to this work.

Yong Liu is with the School of Instrumentation and Optoelectronic Engineering, Beihang University, Beijing, China. (e-mail: yongliu@buaa.edu.cn).

Mengtian Kang was with Beijing Tongren Hospital, Capital Medical University, Beijing, China (e-mail: kangmengtian@163.com).

Shuo Gao is with the School of Instrumentation and Optoelectronic Engineering, Beihang University, Beijing, China. (e-mail: shuo_gao@buaa.edu.cn).

Chi Zhang is with Beijing Tongren Hospital, Capital Medical University, Beijing, China. (e-mail: czhang0426@163.com)

Ying Liu is with the Department of Surgery (Ophthalmology), The University of Melbourne, Melbourne, Australia. (e-mail: 448689563@qq.com)

Shiming Li is with Beijing Tongren Hospital, Capital Medical University, Beijing, China. (e-mail: lishiming81@163.com)

Yue Qi is with Beijing Tongren Hospital, Capital Medical University, Beijing, China. (e-mail: qiyue@126.com)

Arokia Nathan is with Darwin College, University of Cambridge, Cambridge, UK. (e-mail: an299@cam.ac.uk)

Wenjun Xu is with Beijing Tongren Hospital, Capital Medical University, Beijing, China. (e-mail: sallyxuwenjun@163.com)

Chenyu Tang is with the Department of Engineering, University of Cambridge, Cambridge, UK. (e-mail: ct631@cam.ac.uk)

Edoardo Occhipinti is with the Department of Computing, Imperial College London, UKRI Centre for Doctoral Training in AI for Health, London, UK. (e-mail: edoardo.occhipinti16@imperial.ac.uk)

Mayinuer Yusufu is with the Department of Surgery (Ophthalmology), The University of Melbourne, Melbourne, Australia. (e-mail: mayinuer.yusufu@student.unimelb.edu.au)

Ningli Wang is with Beijing Tongren Hospital, Capital Medical University, Beijing, China. (e-mail: wningli@vip.163.com)

Weiling Bai is with Beijing Tongren Hospital, Capital Medical University, Beijing, China. (e-mail: 15811025078@163.com)

Luigi Occhipinti is with the Department of Engineering, University of Cambridge, Cambridge, UK. (e-mail: lgo23@cam.ac.uk)

# Diagnosis of Multiple Fundus Disorders Amidst a Scarcity of Medical Experts Via Self-supervised Machine Learning

Yong Liu, Mengtian Kang, Shuo Gao, Chi Zhang, Ying Liu, Shiming Li, Yue Qi, Arokia Nathan, Wenjun Xu, Chenyu Tang, Edoardo Occhipinti, Mayinuer Yusufu, Ningli Wang, Weiling Bai, and Luigi Occhipinti

***Abstract*— Fundus diseases are major causes of visual impairment and blindness worldwide, especially in underdeveloped regions, where the shortage of ophthalmologists hinders timely diagnosis. AI-assisted fundus image analysis has several advantages, such as high accuracy, reduced workload, and improved accessibility, but it requires a large amount of expert-annotated data to build reliable models. To address this dilemma, we propose a general self-supervised machine learning framework that can handle diverse fundus diseases from unlabeled fundus images. Our method's AUC surpasses existing supervised approaches by 15.7%, and even exceeds performance of a single human expert. Furthermore, our model adapts well to various datasets from different regions, races, and heterogeneous image sources or qualities from multiple cameras or devices. Our method offers a label-free general framework to diagnose fundus diseases, which could potentially benefit telehealth programs for early screening of people at risk of vision loss.**

## I. INTRODUCTION

Visual impairment and blindness pose significant challenges to public health and socio-economic development worldwide. The prevalence of these conditions is high, with an estimated 61 million people expected to suffer from blindness and 474 million from moderate to severe visual impairment by 2030 (1). Visual impairment and blindness carry a substantial economic burden, with a projected global cost of $537.6 billion in potential productivity losses by 2030 (2). Early detection of retinal diseases is crucial in preventing visual loss. Unfortunately, there are only 3.7 ophthalmologists per 100,000 population globally, and this ratio is even lower in Africa and Southeast Asia, at 0.7 and 1.6 respectively, due to the high technical level and extended duration of training required to become professional ophthalmologists (3). The shortage of ophthalmologists makes timely diagnosis of fundus diseases almost impossible, which poses serious health risks to patients.

Recent advances in electronic and computer technology have enabled the use of AI methods based on digital fundus photographs for large-scale screening of various fundus diseases. However, most state-of-the-art methods rely on manual annotation by doctors (4-6), i.e., labeling the fundus images with the correct disease categories or regions of interest. This process is not only time-consuming and labor-intensive, but also prone to human error or inconsistency, especially when the disease types are diverse and numerous. Moreover, it limits the scalability of datasets, as there is a huge gap between the number of available fundus images and the number of annotated images. Another challenge is the diagnosis of multiple fundus diseases with one framework, which can reduce the detection and classification performance of the proposed model. For instance, diabetic retinopathy (DR), glaucoma, and age-related macular degeneration are common fundus diseases that exhibit different pathological features on the fundus (7). Recent studies tend to focus on a single disease or a limited number of diseases (8-9).

To address the problem, this study proposes label-free general framework based on self-supervised machine learning, namely LSVT-Net. Specifically, our model performs feature distillation on a large number of unlabeled fundus images and employs a linear classifier for the detection of different fundus diseases. This approach allows the model to learn useful representations of the input fundus images, which can be used for downstream tasks such as fundus disease detection and classification.

Our model achieves state-of-the-art performance in detecting various fundus diseases, without requiring any disease-specific labels. By developing and validating our approach on public and external validation fundus datasets, our model surpasses existing methods and even exceeds human expert-level performance, with up to 15.7% higher Area Under the Curve (AUC). Moreover, our model can generalize well to different datasets from diverse regions, races, and image sources or qualities from multiple cameras or devices. By providing a label-free general framework, our method enables accurate diagnosis of a wide range of fundus diseases, which could potentially benefit telehealth programs for early detection of people at risk of vision loss.

## II. MATERIALS AND METHODS

### A. Study approval

This study was approved by the Ethics Committee of Beijing Tongren Hospital Capital Medical University, approval ref. No. TRecky2019-025, and adhered to the tenets of the Declaration of Helsinki. Informed consent was not obtained from patients due to the anonymity and retrospective nature of the study.

| Data sets | Age[a], mean(SD) | Men[a], no.(%) | Images | Patients | Referable(%) Images | Referable(%) Patients |
|---|---|---|---|---|---|---|
| **Self-supervised training** | | | | | | |
| EyePACS | N/A | N/A | 88,702 | 44,351 | 23,360(26.3%) | N/A |
| **Total, training** | N/A | N/A | 88,702 | 44,351 | 23,360(26.3%) | N/A |
| **Classifier training** | | | | | | |
| Messidor | 57.6(15.9) | 57.4 | 960 | 699 | 523(54.5%) | N/A |
| Messidor-2 | 59.9(11.7) | 57.4 | 1,395 | 698 | 584(41.8%) | N/A |
| APTOS-2019 | 53.5 | N/A | 2,930 | N/A | 1,486(50.7%) | N/A |
| Ichallenge-AMD | N/A | N/A | 320 | N/A | 71(22.2%) | N/A |
| REFUGE | N/A | N/A | 640 | N/A | 64(10%) | N/A |
| Ichallenge-PM | N/A | N/A | 320 | N/A | 170(53.1%) | N/A |
| GON* | N/A | N/A | 1,535 | 870 | 378(24.6%) | 192 |
| PM* | N/A | N/A | 1,790 | 723 | 163(9.1%) | 96 |
| **Total, training** | N/A | N/A | 9,890 | N/A | 3,439(34.8%) | N/A |
| **Test** | | | | | | |
| Messidor | 57.6(15.9) | 57.4 | 240 | 175 | 131(54.6%) | N/A |
| Messidor-2 | 57.6 | 57.4 | 349 | 174 | 146(41.8%) | N/A |
| APTOS-2019 | 53.5 | N/A | 732 | N/A | 371(50.7%) | N/A |
| Ichallenge-AMD | N/A | N/A | 80 | N/A | 18(22.5%) | N/A |
| REFUGE | N/A | N/A | 160 | N/A | 16(10%) | N/A |
| Ichallenge-PM | N/A | N/A | 80 | N/A | 43(53.8%) | N/A |
| GON* | N/A | N/A | 384 | 218 | 94(24.5%) | 48 |
| PM* | N/A | N/A | 447 | 181 | 44(9.8%) | 24 |
| **Total, test** | N/A | N/A | 2,472 | N/A | 863(34.9%) | N/A |

[a]Age and gender information cannot be obtained are marked as "N/A".

[b]Private collected dataset are marked with *

## B. Data sets and labeling

To develop LSVT-Net, we collected color fundus photographs from six publicly available datasets and two self-collected external datasets. Publicly available datasets with a total of 96,908 images included EyePACS, APTOS-2019, Messidor and Messidor-2 for DR, AMD for age-related macular degeneration, REFUGE for glaucoma and PM for pathological myopia. Self-collected external datasets with 1,919 images for glaucoma and 2,237 images for pathologic myopia were collected from Beijing Tongren Hospital using different types of non-mydriatic fundus cameras (e.g., Topcon, Canon, Zeiss). The detailed characteristics of all datasets are summarized in Table 1.

During the image labeling process, all fundus images were first de-identified to remove any patient-related information. Images with poor quality (optic disc or macula obscured) or incorrect field definition (non-macula centered fundus photographs) were excluded manually. Five junior ophthalmologists with at least 3 years of clinical experience and three senior retinal specialists with more than 10 years of clinical experience were recruited for image labeling. Images were first labeled by the junior ophthalmologists and then confirmed by the senior retinal specialist. Specifically, we used a hierarchical two-tier grading process involving 5 phase I and 3 phase II graders. Phase I graders underwent training and evaluation to achieve at least 95% accuracy on a quiz consisting of 1,000 fundus images with various retinal diseases. Phase II graders were ophthalmologists who individually reviewed every image classified by phase I graders.

To ensure consistency among phase II graders, we randomly selected 20% of images and had them reviewed by three senior retinal specialists. The final tier of five ophthalmologists independently reviewed and verified the true labels for each image. In case of disagreement, we resolved it by expert consensus. About 5% of the study participants were excluded due to poor photographic quality, unreadable images or missing clinical diagnosis. After establishing the consensus diagnoses, we transferred the images to the AI team to develop a deep-learning algorithm for image-based classification.

### C. Criteria of glaucoma and pathologic myopia

The diagnosis of glaucoma was in accordance with the criteria of previous population-based studies (13). Glaucomatous optic neuropathy (GON) was defined by the presence of a vertical cup-to-disc ratio of 0.7 or greater, a retinal nerve fiber layer defect or localized notching, an optic disc rim width of 0.1disc diameter or less, and/or a disc splinter hemorrhage. An eye would be labeled as referable GON (either suspected or confirmed) if any of the above criteria were met.

Pathological myopia was diagnosed according to the Meta-analysis for Pathological Myopia (META-PM) classification. Briefly, this classification defines five categories of myopic maculopathy and three "plus" lesions. The five categories of myopic maculopathy include "no myopic retinal lesions" (category 0), "tessellated fundus (TF) only" (category 1), "diffuse chorioretinal atrophy" (category 2), "patchy chorioretinal atrophy" (category 3) and "macular atrophy" (category 4). Three "plus" lesions included lacquer cracks, myopic choroidal neovascularization and Fuchs' spot. Pathological myopia was defined as category 2 or greater or the presence of any plus lesion.

### D. Outcomes

The primary output of the LSVT-Net is the prediction of the grade of lesions for each fundus disease, including DR, glaucoma, age-related macular degeneration, and pathological myopia. Among these, diabetic retinopathy is divided into 5 classes: no evidence of retinopathy, mild retinopathy, moderate retinopathy, severe retinopathy and proliferative retinopathy. Glaucoma, age-related macular degeneration and pathologic myopia are divided into 2 classes: not apparent and apparent.

### E. Algorithm development

The LSVT-Net is based on the architecture of Deit-B, a transformer-based model. It differs from convolutional neural networks (CNNs) in that the attention mechanism replaces all convolutional operations. The transformer attention is formed with basic scaled dot-product attention. It is calculated through the matrices of queries ($Q$), keys ($K$) and values ($V$), as described in Eq. (1):

$$\text{attention}(Q, K, V) = \text{softmax}\left(\frac{QK^T}{\sqrt{d_k}}\right)V$$

(1)

where $d_k$ is the dimension of the queries ($q$) and keys ($k$) vectors. In the following we will denote $d_v$ as the dimension of the values vector ($v$). $W^Q$, $W^K$ and $W^V$ denote projection matrices used to generate $Q$, $K$ and $V$ matrices. $W^O$ is projection matrix for the multi-head output.

In detailed procedure, alignment scores are first calculated by multiplying queries matrix $Q(m \times d_k)$ and keys matrix $K(n \times d_k)$, resulting in a matrix sized $m \times n$, as described in Eq. (2),

$$QK^T = \begin{bmatrix} e_{11} & e_{12} & \cdots & e_{1n} \\ e_{21} & e_{22} & \cdots & e_{2n} \\ \vdots & \vdots & \ddots & \vdots \\ e_{m1} & e_{m2} & \cdots & e_{mn} \end{bmatrix}$$

(2)

then the alignment scores are scaled by $\frac{1}{\sqrt{d_k}}$, as described in Eq. (3),

$$\frac{QK^T}{\sqrt{d_k}} = \begin{bmatrix} \frac{e_{11}}{\sqrt{d_k}} & \frac{e_{12}}{\sqrt{d_k}} & \cdots & \frac{e_{1n}}{\sqrt{d_k}} \\ \frac{e_{21}}{\sqrt{d_k}} & \frac{e_{22}}{\sqrt{d_k}} & \cdots & \frac{e_{2n}}{\sqrt{d_k}} \\ \vdots & \vdots & \ddots & \vdots \\ \frac{e_{m1}}{\sqrt{d_k}} & \frac{e_{m2}}{\sqrt{d_k}} & \cdots & \frac{e_{mn}}{\sqrt{d_k}} \end{bmatrix}$$

(3)

then weights are obtained by softmax operation as described in Eq. (4),

$$\text{Softmax}\left(\frac{QK^T}{\sqrt{d_k}}\right) = \begin{bmatrix} \text{Softmax}(\frac{e_{11}}{\sqrt{d_k}} & \frac{e_{12}}{\sqrt{d_k}} & \cdots & \frac{e_{1n}}{\sqrt{d_k}}) \\ \text{Softmax}(\frac{e_{21}}{\sqrt{d_k}} & \frac{e_{22}}{\sqrt{d_k}} & \cdots & \frac{e_{2n}}{\sqrt{d_k}}) \\ \vdots & \vdots & \ddots & \vdots \\ \text{Softmax}(\frac{e_{m1}}{\sqrt{d_k}} & \frac{e_{m2}}{\sqrt{d_k}} & \cdots & \frac{e_{mn}}{\sqrt{d_k}}) \end{bmatrix}$$

(4)

the attention is obtained through weights multiply values matrix $V(n \times d_v)$, as described in Eq. (5),

$$\text{Softmax}\left(\frac{QK^T}{\sqrt{d_k}}\right) \bullet V =$$
$$\begin{bmatrix} \text{Softmax}(\frac{e_{11}}{\sqrt{d_k}} & \frac{e_{12}}{\sqrt{d_k}} & \cdots & \frac{e_{1n}}{\sqrt{d_k}}) \\ \text{Softmax}(\frac{e_{21}}{\sqrt{d_k}} & \frac{e_{22}}{\sqrt{d_k}} & \cdots & \frac{e_{2n}}{\sqrt{d_k}}) \\ \vdots & \vdots & \ddots & \vdots \\ \text{Softmax}(\frac{e_{m1}}{\sqrt{d_k}} & \frac{e_{m2}}{\sqrt{d_k}} & \cdots & \frac{e_{mn}}{\sqrt{d_k}}) \end{bmatrix} \bullet \begin{bmatrix} v_{11} & v_{12} & \cdots & v_{1d_v} \\ v_{21} & v_{22} & \cdots & v_{2d_v} \\ \vdots & \vdots & \ddots & \vdots \\ v_{n1} & v_{n2} & \cdots & v_{mn} \end{bmatrix}$$

(5)

For the multi-head attention, there are $h$ sets of $Q$, $K$, $V$, $W^Q$, $W^K$ and $W^V$, producing $h$ outputs. The multi-head attention function is described in Eq. (6),

| Year | Publication | Disease | Disease Type | Requiring labeling | AUC | Dataset Scalability |
|------|-------------|---------|--------------|--------------------|-----|---------------------|
| 2021 | Nature Communication (8) | DR | 1 | Yes | 0.916-0.970 | Not mentioned |
| 2022 | Nature Biomedical Engineering (15) | DR | 1 | Yes | 0.970 | Not mentioned |
| 2021 | Nature Biomedical Engineering (5) | Chronic kidney disease, Type 2 diabetes | 2 | Yes | 0.930 | Not mentioned |
| 2018 | Nature Medicine (6) | DR, AMD, GON, | 3 | Yes | 0.958-0.993 | Device-independent |
| 2022 | Nature Machine Intelligence (4) | DR, AMD, GON, Retinal occlusion | 4 | Yes | 0.919-0.985 | Multiethnic populations |
| **2023** | **This work** | **DR, AMD, PM, GON** | **4** | **No** | **0.989-0.993** | **High scalability, can handle different quality and source images, multiethnic populations** |

ᵃ DR: Diabetic retinopathy; AMD: Age-related macular degeneration; GON: Glaucoma; PM: Pathological myopia.

each head represents an attention function calculated through its own projection matrices, as described in Eq. (7),

$$\text{head}_i = \text{attention}(QW_i^Q, KW_i^K, VW_i^V) \tag{7}$$

The LSVT-Net was implemented in two stages. The flowchart of the algorithm and the model construction is shown in Fig. 1. In the first step, the dataset was used with EyePACS and the labels corresponding to the fundus images were removed. The unlabeled large-scale fundus images were used to complete the self-supervised learning and output a high-dimensional vector containing semantic information about the fundus. A total of 88,703 fundus images from EyePACS were first normalized and standardized to a $256 \times 256$ size. Global cropping and local cropping were performed to transform each fundus image into a global replica and a local replica. Two sets of fundus images underwent random horizontal flip and random grayscale transformation operations to enter the global Deit-B model and local Deit-B model, respectively, as described in Eq. (8),

$$X_g = \bigcup_{x \in X} \{F_g(x, \varphi_i), i = 1, \cdots, n_g\}$$
$$X_l = \bigcup_{x \in X} \{F_l(x, \varphi_i), i = 1, \cdots, n_l\} \tag{8}$$

where $X_{g,l}$ represents global or local crop sets, $g$ for global and $l$ for local. The crop numbers are predefined as $n_{g,l}$. $F_{g,l}(x, \varphi_i)$ are the global or local crop operations, and $\varphi_i$ the subsequent transformations, which includes a random crop and a random flip.

The global and local Deit-B models have identical model structures, but the gradient function is eliminated from the global Deit-B model. The global and local models output high-dimensional feature vectors, and the cross-entropy loss between them is calculated after softmax operation, as described in Eq. (9),

$$P_t^i(x_g, \theta_t) = \frac{\exp\left(\frac{G_t^i(x_g, \theta_t)}{\tau_t}\right)}{\sum_{k=1}^{K} \exp\left(\frac{G_t^i(x_g, \theta_t)}{\tau_t}\right)}$$
$$H(\alpha, \beta) = -\alpha \log \beta \tag{9}$$

where $G_t^i(x_g, \theta_t)$ means i-th class output among the K classes. $P_t^i(x_g, \theta_t)$ represents i-th class output probability calculated through softmax operation with a temperature $\tau$, $x_g$ for input global crop, $\theta_t$ for teacher's weights. $H(\alpha, \beta)$ is the cross-entropy loss.

The loss updates the weight parameters of the local model by back-propagation through Eq. (10). The gradient of the global model is cancelled and therefore not updated by back propagation. It is updated by an exponential moving average of the historical weights of the local model.

$$\min_{\theta_s} \sum_{x_g^i \in X_g, x_l^i \in X_l} H\left(P_t(x_g^i, \theta_t), P_s(x_l^i, \theta_s)\right)$$
$$\text{subject to } \theta_t^{epoch} = \lambda\ \theta_t^{epoch-1} + (1-\lambda)\theta_s^{epoch-1} \tag{10}$$

In the second stage, the dataset was used to determine the grade of fundus disease using datasets of Messidor-2, AMD, REFUGE, PM, etc. The global model obtained by self-supervised learning was used to compute its high-dimensional feature vectors for fundus images and construct a linear classifier consisting of a single-layer neural network for fundus disease classification. We used 60% of the fundus image dataset as the training set for training the linear classifier, the other 20% as a validation set for hyperparameter tuning of the linear classifier, and the remaining 20% as a test set to verify the performance evaluation of the disease classification. Partitioning was performed evenly according to class distribution, ensuring equal distribution of fundus image data for different classes.

### F. Evaluation Protocol

Evaluation metrics for the performance of the deep learning system included the area under the receiver operating characteristic curve (AUC), accuracy, and F1 score, with AUC as the main metric because of its merits in providing a comprehensive assessment of sensitivity and specificity. In the case of multiple classes of lesions in DR, the AUC values were calculated separately for each of the two categories and then the average value was taken as the final result.

## III. EXPERIMENTS AND RESULTS

### A. Data characteristics

We developed the self-supervised learning part of the LSVT-Net using 88,702 unlabeled fundus images, and the linear classifier for disease diagnosis using 9,890 fundus images of DR, glaucoma, age-related macular degeneration, and pathological myopia. The performance of the LSVT-Net was then evaluated using a total of 1,641 external validation images collected from the Beijing Tongren Hospital. These images included participants who underwent an annual health check. The basic characteristics, demographics, and disease class distribution of the training dataset and the external validation datasets are shown in Table 1.

### B. Performance comparison

As shown in Fig. 2a, we evaluated our model on six public fundus disease datasets (Messidor, Messidor-2 and APTOS for DR, Ichallenge-AMD for age-related macular degeneration, Ichallenge-PM for pathological myopia and REFUGE for glaucoma), and two external datasets. LSVT-Net achieved an average AUC of 98.2% across all datasets, with a maximum of 99.3%. In addition, LSVT-Net achieved an AUC of 0.998 for up to 4 fundus diseases without image annotation (Table 2).

As for public datasets validation, our model outperformed previous state-of-the-art methods based on supervised learning and other approaches (Table 3). The performance improvement ranged from 0.7% to 15.7% in terms of AUC, demonstrating the effectiveness and robustness of our model.

### C. Ablation study

To measure the effectiveness of self-supervised learning, we used a portion of the training set to construct the linear classifier. This was done to verify that the high-dimensional feature vector obtained from the self-supervised model has broad representational significance. We reduced the training data set used to train the disease diagnosis classifier to 6.5%, 10%, 25%, 50%, and 75%, respectively, and compared the results with the original 100% training data.

The experimental results, as shown in Fig. 2b, indicated that the classification performance for each fundus disease would improve significantly with the increase in the percentage of training data. Notably, the classification performance AUC reached 0.907 on the APTOS dataset using only 6.5% of the labeled fundus image data. We also examined how the size of the self-supervised training dataset affects the classification performance for each fundus disease. We found that using more unlabelled fundus images for self-supervised learning leads to better results, and there is no sign of diminishing returns. This suggests that LSVT-Net can benefit from further scaling up the self-supervised training data.

In the process of tuning the hyperparameters of the linear classifier using the validation set, we conducted experiments on the effect of the dropout rate of the linear classifier using dropout rates of 0, 0.1, 0.2, and 0.5. In the validation set, the recognition performance for each fundus disease tended to increase and then decrease as the dropout rate increased. The highest overall performance was achieved at the dropout rate of 0.1. There are two ways to optimize the training of linear classifiers: the frozen approach and the end-to-end approach.

The frozen approach means freezing the self-supervised model and only optimizing the weights of the linear classifier. The end-to-end approach represents the joint optimization of the self-supervised model and the linear classifier in an end-to-end manner. The experimental results on Messidor and Messidor-2 datasets, as shown in Fig. 2b, showed that the frozen approach performed better than the end-to-end approach.

On the APTOS dataset, the end-to-end approach outperformed the frozen approach. The possible explanation for this phenomenon is that the end-to-end approach needs to optimize more model weights and thus requires a larger amount of data to support it, while overfitting occurs when the amount of data is too small. The frozen approach, on the other hand, only needs to optimize the weight parameters of the linear classifier and tends to perform better on small datasets.

Fig. 2c illustrates the normalized confusion matrices for eight datasets, five of which have only 2 disease categories. It was evident that, e.g. in the Messidor and Messidor 2 datasets, classifying mild fundus diseases was more challenging for LSVT-Net. In the APTOS dataset, it was more difficult to classify the severe retinopathy class. For the other two-class diseases, it was harder to identify the type of the disease.

To validate the effectiveness of our self-supervised learning model, we used the linear separability of the output high-dimensional vector to describe it. Strong linear separability indicated that our self-supervised model could learn the rich semantic information of the fundus without the need for labeling information. Therefore, we performed a t-SNE dimensionality reduction visualization of the output high-dimensional vectors (9). t-SNE is a statistical method for visualizing high-dimensional data in a low-dimensional space of two or three dimensions. It transforms the similarities between data points into joint probabilities and minimizes the divergence between the original and the embedded data distributions. t-SNE is useful for exploring the structure and patterns of complex datasets. The dimensionality reduction visualization for each dataset is shown in Fig. 3a. It can be seen that the different classes of each fundus disease were clustered into unique shape distributions in the 2D plane, especially in the binary dataset where the normal and diseased

TABLE III

COMPARISON WITH RELATED STUDIES ON PUBLIC DATASETS

| Training dataset | Year | Benchmark | Learning algorithm | Approach | AUC (%) | ACC (%) | F$_1$ score |
|---|---|---|---|---|---|---|---|
| Messidor 1200 images DR:4 classed | 2021 | A. K. Gangwar, et al.(*16*) | Supervised | Inception | / | 72.3 | / |
| | 2019 | X. Li, et al.(*17*) | Supervised | ResNet50 | 96.3 | 85.1 | / |
| | 2020 | G. Saxena, et al.(*18*) | Supervised | Inception-v3 | 95.8 | / | / |
| | 2021 | B. B. Marc, et al.(*19*) | Supervised | DCNN | 95.9 | 94.8 | / |
| | 2022 | A. S. Hervella, et al.(*20*) | Unsupervised | VGG | 94.2 | 79.4 | / |
| | **2023** | **This work** | **Self-supervised** | **LSVT-Net** | **96.1** | **88.8** | **86.7** |
| Messidor-2 1,748 images DR:5 classes | 2019 | R. Pires, et al.(*21*) | Supervised | VGG-16 | 96.9 | / | / |
| | 2021 | G. Kumar, et al.(*22*) | Supervised | DRISTI | / | 84.1 | / |
| | 2021 | H. Ríos, et al.(*23*) | Supervised | DenseNet121 | 94.8 | / | / |
| | 2022 | M. R. Islam, et al.(*24*) | Supervised | Supervised | 87.3 | 74.2 | / |
| | 2022 | J. M. Arrieta Ramos, et al.(*25*) | Self-supervised | DenseNet161 | 89.0 | / | / |
| | **2023** | **This work** | **Self-supervised** | **LSVT-Net** | **96.6** | **92.7** | **90.5** |
| APTOS-2019 3,662 images DR:5 classes | 2020 | M. Chetoui, et al.(*26*) | Supervised | EfficientNet | 96.6 | / | / |
| | 2019 | N. Khalifa, et al.(*27*) | Supervised | AlexNet | / | 97.9 | 95.8 |
| | 2020 | S. Taufiqurrahman, et al.(*28*) | Supervised | MobileNet | 93.0 | / | / |
| | 2019 | S. H. Kassani, et al(*12*) | Supervised | Xception | 91.8 | 83.1 | / |
| | 2022 | S. Duan, et al.(*29*) | Semi-supervised | GACNN | / | 89.5 | / |
| | **2023** | **This work** | **Self-supervised** | **LSVT-Net** | **98.1** | **94.0** | **93.0** |
| Ichallenge-AMD 400 images AMD:2 classes | 2022 | R. Chakraborty, et al.(*30*) | Supervised | DCNN | / | 89.8 | / |
| | 2020 | X. Li, et al.(*13*) | Self-supervised | CycleGAN | 83.2 | 89.4 | 83.7 |
| | 2021 | Z. Zhang, et al.(*11*) | Unsupervised | CycleGAN | 86.2 | 87.3 | 73.2 |
| | 2021 | X. Li, et al.(*31*) | Unsupervised | ResNet18 | 87.0 | 90.6 | 86.4 |
| | 2022 | A. Chen, et al.(*32*) | Supervised | PRGAN | / | 95.8 | / |
| | **2023** | **This work** | **Self-supervised** | **LSVT-Net** | **98.9** | **99.1** | **91.2** |
| REFUGE 400 images GON:2 classes | 2020 | Y. Sun, et al.(*33*) | Unsupervised | UDAGAN | / | 89.8 | / |
| | 2021 | B. P. Yap, et al.(*34*) | Semi-supervised | ResNet-50 | 98.8 | / | / |
| | 2020 | S. Sreng, et al.(*35*) | Supervised | DenseNet | 94.6 | 93.0 | / |
| | 2018 | A. Chakravarty, et al.(*36*) | Supervised | U-Net | 96.0 | / | / |
| | 2021 | S. S. Ganesh, et al.(*37*) | Supervised | ResNet | / | 99.5 | 97.6 |
| | **2023** | **This work** | **Self-supervised** | **LSVT-Net** | **99.0** | **99.6** | **89.0** |
| Ichallenge-PM 400 images PM:2 classes | 2021 | J. Cui, et al.(*38*) | Supervised | VGG-16 | / | 95.0 | / |
| | 2020 | X. Li, et al.(*13*) | Self-supervised | CycleGAN | 98.6 | 98.7 | 98.6 |
| | 2021 | Z. Zhang, et al.(*11*) | Unsupervised | CycleGAN | 100 | 100 | 100 |
| | 2021 | X. Li, et al.(*31*) | Unsupervised | ResNet18 | 99.1 | 99.2 | 99.2 |
| | 2019 | R. Zhang, et al.(*39*) | Supervised | GAN-CNN | / | 98.5 | / |
| | **2023** | **This work** | **Self-supervised** | **LSVT-Net** | **99.2** | **99.3** | **98.0** |

[a] Results not provided are marked with "/".

In addition, to compare the discriminability of different classes within each fundus disease dataset, we also analyzed the discriminability between each fundus disease. Each data point represents the average of all sample points for a fundus disease class in a dataset, as shown in Fig. 3b. It is noteworthy that the two data sets associated with pathological myopia were clustered together in the two-dimensional plane, whereas the data points for age-related macular degeneration were distributed separately. A number of normal or mild data points were concentrated near the center of the two-dimensional plane. From a clinical perspective, pathological myopia, age-related macular degeneration, and glaucoma affect different regions of the fundus: the choroid, macular area, and optic disc, respectively. The t-SNE visualization shows that these three diseases are clearly separated, confirming the ability of LSVT-Net to extract features that focus on distinct parts of the fundus images. The central cluster of the visualization comprises data points without disease or with DR class 1. Ideally, data points without disease should form a distinct group. However, the inclusion of DR class 1 in this cluster suggests that LSVT-Net struggles to differentiate between normal and mildly affected images. The peripheral distribution of DR, glaucoma, age-related macular degeneration, and pathological myopia reflected the increasing accuracy of LSVT-Net in identifying different diseases. This is consistent with the clinical diagnosis of disease based on fundus photography, which becomes easier from the center to the edge.

Fig. 4 compares the attention maps from the self-supervised learning model and the lesion maps labeled by the ophthalmologists. Ophthalmologists usually draw lesion maps based on their clinical expertise, representing known pathology or lesion areas in fundus images. Comparing attention maps with lesion maps helps to evaluate the model's ability to recognize and emphasize areas consistent with known disease similarity with the maps marked by experts for four different markers. The model's attention maps show a high degree of diseases. This indicates that the self-supervised learning model can effectively capture the known fundus biomarkers associated with these diseases. Among the areas of interest that the deep learning model focuses on, some are fundus biomarkers, while others are unknown. This finding is interesting, as it means that the model may identify new potential biomarkers or features in fundus images that have not been previously recognized by experts. The identification of previously unknown areas of interest by the deep learning model is of great significance for medical imaging research, as researchers can further investigate these areas to determine whether they are indeed indicators of disease, or whether they represent new features that can be used to improve diagnostic accuracy or to facilitate disease detection and understanding, thereby improving patient prognosis.

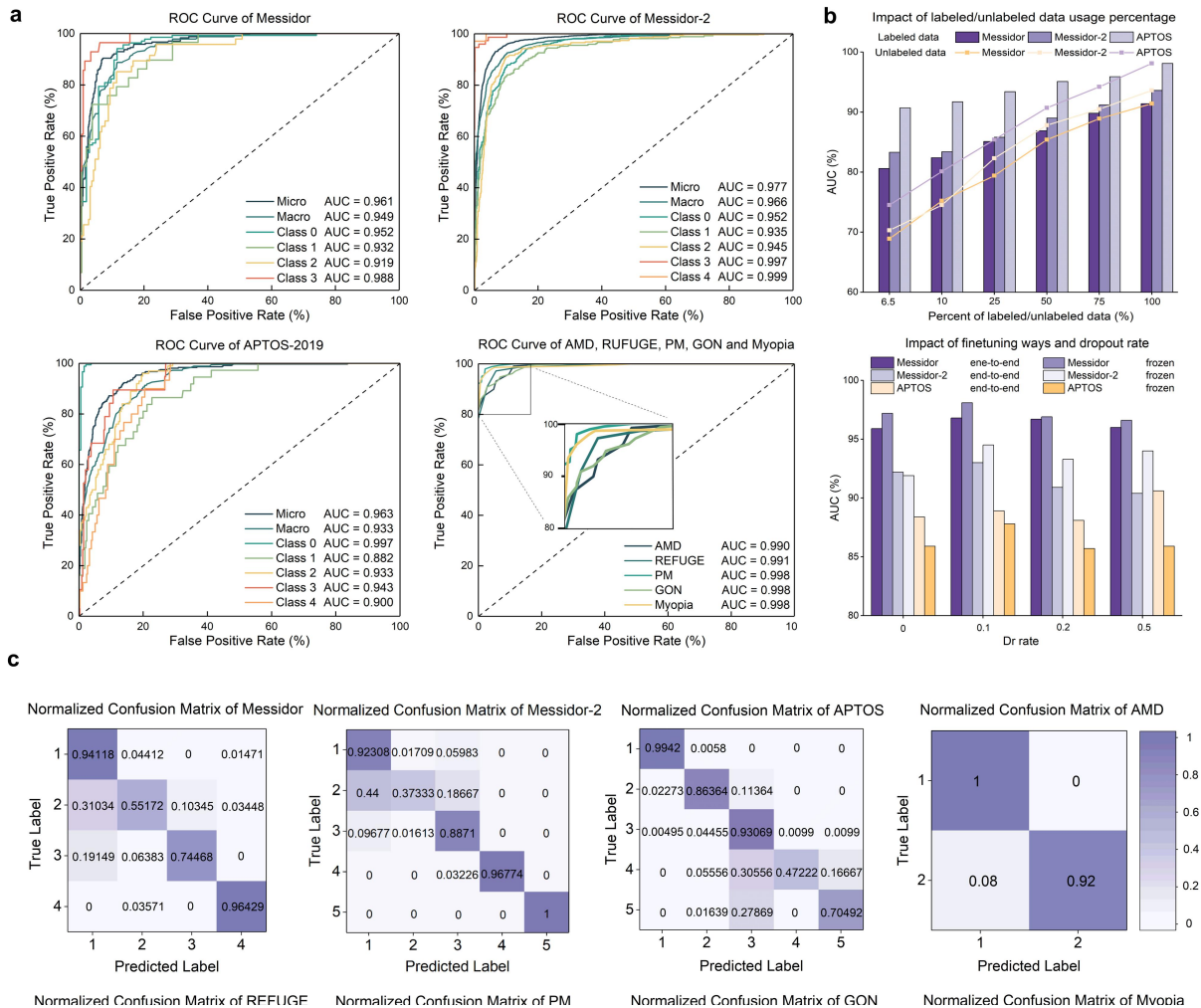

**Fig. 2. Experimental results in terms of ROC curve, hyperparameter adjustment and normalized confusion matrix.**
**(a)** AUC of all datasets including 6 public validation datasets and 2 external validation datasets. For the multiclassification dataset, we also plotted the ROC curve for each category separately. **(b)** Impact of labeled and unlabeled data usage percentage and comparisons of different finetuning ways on each dataset. The richness and robustness of the features learned by the self-supervised learning model were verified by adjusting the percentage of labeled data used. The optimal performance point of the model was identified by adjusting hyper-parameters such as finetuning approach and dropout rate. **(c)** Normalized confusion matrix of 6 public validation datasets and 2 external validation datasets. The difference in performance of the model for different fundus disease datasets on different categories was reflected by the confusion matrix.

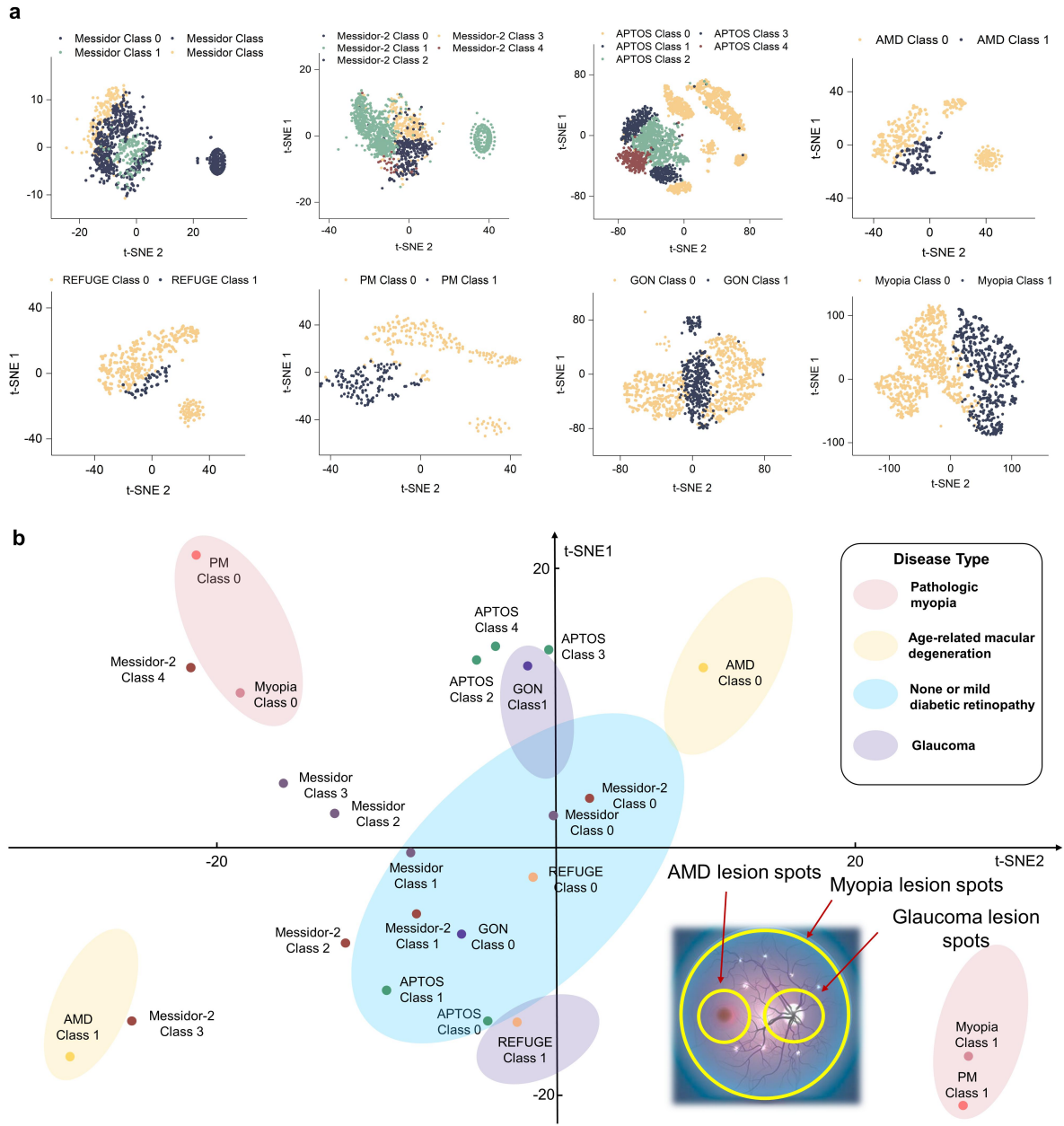

**Fig. 3. T-SNE Visualization of intra-fundus and inter-fundus disease. (a)** t-SNE dimensionality reduction visualization of intra-fundus disease. The t-SNE visualization revealed that the data points of various fundus disease datasets exhibited a clustered distribution in the two-dimensional plane. **(b)** t-SNE dimensionality reduction visualization of inter-fundus disease. Each data point in the graph represents the average of all data points for the disease class to which it belongs. From a medical perspective, it is notable that AMD, PM and GON, which have the lesion points in different locations, showed similar clustering or separation properties in the two-dimensional plane.

## IV. DISCUSSION

To the best of our knowledge, this is the first report to present a label-free self-supervised learning system capable of predicting lesion grade in DR, glaucoma, age-related macular degeneration, and pathological myopia. LSVT-Net leverages unlabeled fundus images to learn rich and robust representations that can be transferred to diagnosis tasks for different fundus diseases. We validated it on eight validation datasets, including six public datasets, such as Messidor-2 and AMD, and two external datasets comprised of glaucoma and pathologic myopia datasets collected from Beijing Tongren Hospital. On both types of datasets, LSVT-Net achieved high performance for fundus lesion classification, demonstrating its ability to handle different quality images from various fundus cameras and diagnose fundus diseases across different races. Furthermore, LSVT-Net can generalize to other fundus diseases without requiring a large number of labeled data, which is a significant advantage over existing methods.

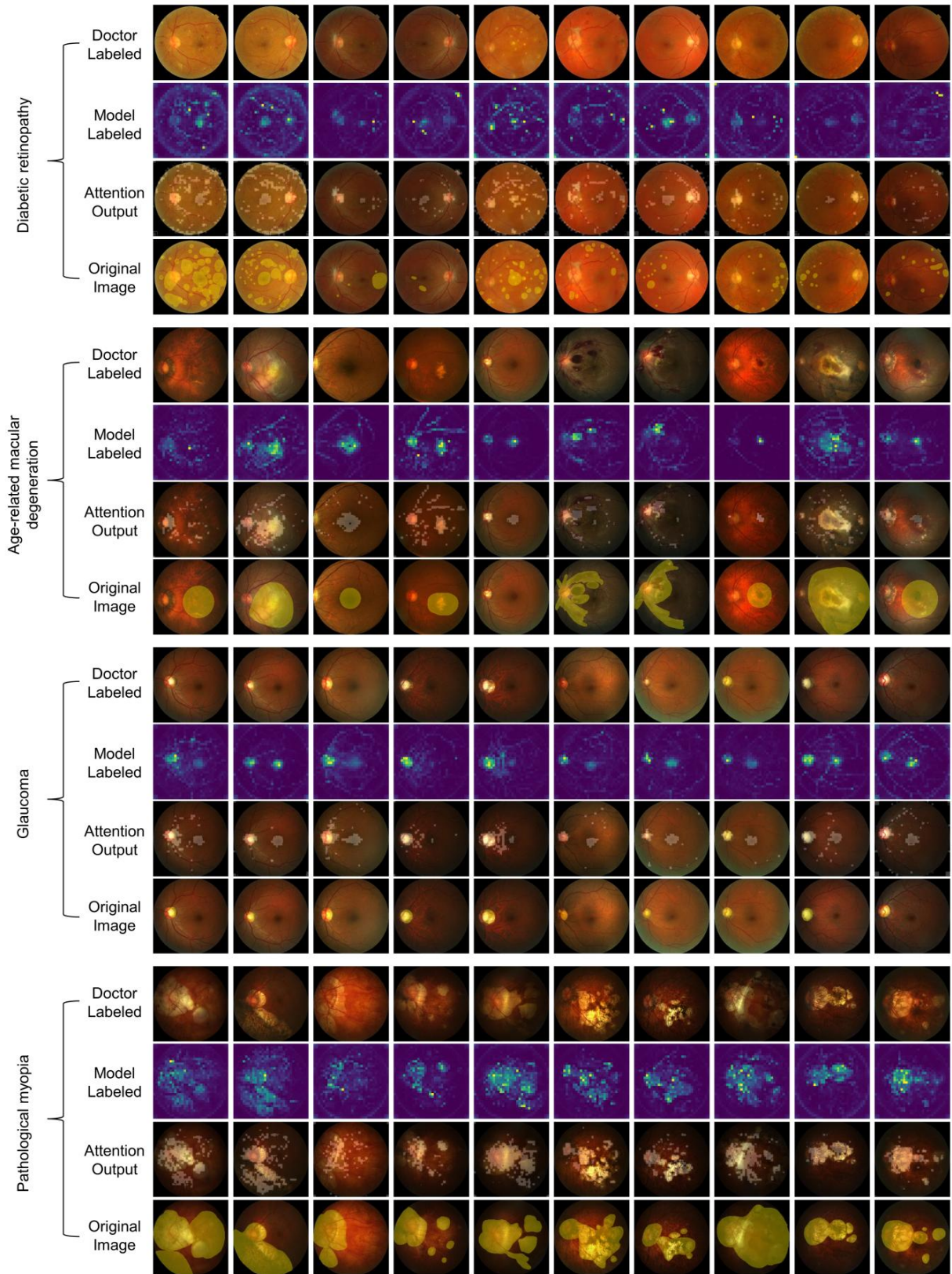

**Fig. 4. Comparison between the attention map produced by the self-supervised learning model and ophthalmologists.** The heat map output was generated from the final attention layer of the self-supervised model, and the physician annotated maps were manually annotated by an ophthalmologist with three years of clinical experience in the field.

Previous studies have explored fundus disease classification using different approaches, such as the classification of DR grades using a contrast learning approach based on the ResNet50 architecture (10), or the detection of age-related macular degeneration and pathological myopia using a feature-based learning approach based on cycle generative adversarial networks (11). However, these studies have some limitations, such as only focusing on one or two fundus diseases, poorer detection performance compared to supervised methods, and lack of sufficient external validation sets.

Our study presents several improvements over previous work. First, we proposed a sophisticated self-supervised learning framework that used a large number of unlabeled fundus images for semantic learning, followed by a classification system for grading multiple fundus diseases. The training process of this method does not require the labeling of fundus images and thus can be easily and sustainably obtained from ophthalmic medical institutions, without requiring ophthalmologists to label fundus images for recognition in supervised approaches. By significantly increasing the size of the unlabeled dataset, the self-supervised label-free approach can improve the performance of fundus disease recognition tasks. Second, our system learns broad fundus semantic information in the self-supervised learning phase and showcases high performance in various fundus disease lesion classification tasks. Third, we validated our system on several validation datasets, including six publicly available validation datasets and two self-collected external validation datasets. Despite the differences in patient populations, camera parameters, incidence rates and grading schemes, our system maintained high predictive performance on all validation datasets, surpassing a single human expert's performance. In addition, our self-supervised learning system uses unlabeled fundus images, which can be continuously provided by ophthalmic providers and are therefore continuously evolving.

In addition to validating the experimental results, we conducted additional analysis to better understand the reasons behind the experimental results. First, we examined the performance of the deep learning system when only a fraction of the training data was retained. For example, using only 6.5% of the APTOS training data, we were able to achieve a fairly high recognition performance of 0.907 for AUC. Second, we validated the parameters of the linear classifier for fundus disease recognition and the effect of system fine-tuning to further improve the performance of fundus disease classification. Third, heat map analysis showed that the self-supervised learning model was able to identify biomarkers and lesion regions in the fundus, suggesting that the performance of the linear classifier was based on a well-trained self-supervised model. Fourth, the dimensionality reduction effect of t-SNE showed that the output vector of our self-supervised model contained rich information, both within and between the different classes of individual fundus diseases as well as amongst different fundus diseases.

One of the applications of deep learning systems for fundus disease detection is the provision of technical support for large-scale screening studies. Conventional screening methods rely on a bespoke camera and diagnosis by an ophthalmologist specialized in common fundus diseases such as DR. Deep learning systems can provide patients with a risk assessment for self-screening of fundus diseases, enabling high-risk patients to seek timely diagnosis by professional ophthalmologists, while low-risk patients can avoid unnecessary use of ophthalmology consultation resources.

In developing areas with limited medical resources, deep learning systems can allocate limited medical resources more efficiently and build a better screening system to prevent and manage fundus diseases. Deep learning systems can also provide pre-diagnostic results and preliminary evidence for ophthalmologists, thereby reducing their workload. Feedback from ophthalmologists can help improve the performance of deep learning system through iterative optimization. Therefore, deep learning systems possess the potential to provide reliable diagnosis for a wide range of fundus diseases and to form a highly efficient screening system.

Our study has some limitations. First, this deep learning system was trained for four major retinal diseases and healthy controls, but it did not cover all retinal diseases and conditions. Future studies should extend its coverage to other retinal diseases. Second, although the deep learning system achieved high accuracy on public age-related macular degeneration and DR datasets, it has not been validated on external age-related macular degeneration and DR datasets. Validation with such datasets and datasets from multiple hospitals is necessary to confirm the generalizability of the system.

## V. CONCLUSION

In this study, we presented LSVT-Net, a label-free self-supervised vision transformer network for fundus disease diagnosis. Our model can detect multiple types of fundus diseases without requiring manual annotations by medical doctors. We evaluated our model on four public and two external fundus disease datasets and achieved state-of-the-art performance on all of them. Moreover, it can generalize well to unseen datasets from different regions, races, and image sources or qualities, demonstrating its robustness and adaptability to real-world scenarios. Our model has the potential to facilitate large-scale screening of fundus diseases and empower the prevention of visual impairment and blindness.